\documentclass[a4paper]{PoS}
\usepackage{amsmath,amssymb}

\title{Nuclear correlations and modifications of the nucleon-nucleon potential due to the QCD critical mode}

\ShortTitle{Nuclear correlations and modifications of the $NN$ potential due to the QCD critical mode}

\author{\speaker{Juan M. Torres-Rincon} and Edward Shuryak\\
       Department of Physics and Astronomy, Stony Brook University,
Stony Brook NY 11794-3800, USA\\
        E-mail: \email{jm.torresrincon@gmail.com}, \email{Edward.Shuryak@stonybrook.edu}}

\abstract{The scalar-isoscalar mode of QCD becomes lighter/nearly massless close to the chiral transition/second-order critical point. From nuclear physics we know that this mode is the  main  responsible  for  the  attractive  part  of  the  nucleon-nucleon  potential  at inter-particle distances of 1-2 fm. Therefore one expects that close to the critical point there is a long-range strong attraction among nucleons. Using a Walecka-Serot model for  the  NN  potential  we  study  the  effects  of  the  critical  point  in  a  finite  system  of nucleons  and  mesons  by  solving  classical  Molecular Dynamics+Langevin  equations for  the  freeze-out  conditions  of  heavy-ion  collisions.  Going  beyond  the  mean-field approximation  allows  us  to  account  for  strong  nucleon  correlations  in  the  time evolution,  leading  to  baryon  clustering.  We  observe  that  light  cluster  formation, together  with  an  enhancement  of  higher-order  cumulants  of  the  proton  distribution can signal the presence of the critical point.}

\FullConference{Corfu Summer Institute 2018 "School and Workshops on Elementary Particle Physics and Gravity"\\
		(CORFU2018)\\
		31 August - 28 September, 2018\\
		Corfu, Greece}

\begin{document}

\section{Motivation}

In the present and past editions of the CPOD conference, have been reported many ways to access the signatures of a possible QCD critical point (CP) by performing heavy-ion collisions at different collision energies~\cite{Stephanov:1998dy,Stephanov:1999zu}. The common origin of these proposals are the peculiarities of the second-order phase transition happening at the critical point, and the probability distribution function of the critical mode, the $\sigma$ field~\cite{Stephanov:2004wx}. In an ideal system (static and infinite) the critical region is dominated by the large fluctuations of $\sigma$ and their correlations. Because this field should couple to the baryon number, it is a potentially good starting point to look for empirical signs of the critical behavior~\cite{Stephanov:2008qz}.

A well-known proposal~\cite{Stephanov:2011pb} is to look at high-order moments of the net-proton distribution, and its related cumulants like the (scaled) skewness and kurtosis. These observables can be extracted from protons and antiprotons detected in the experiment within some particular kinematic cut. When plotted as a function of the collision energy, theoretical predictions indicate that a nonmonotonous behavior should be expected for energies close to the critical region. Preliminary results from STAR collaboration~\cite{Luo:2015ewa} in the context of the Beam Energy Scan (BES) program provided an interesting evidence of this behavior for protons and antiprotons with $p_\perp \in (0.4,2.0)$ GeV at midrapidity ($|y|<0.5$) in the most central collisions.

These cumulants will be re-evaluated at even smaller energies with more statistics at the second phase of the BES and FXT programs of RHIC. In addition, other detectors will also dedicate efforts to study the physics of the critical point like the CBM experiment at FAIR, MPG experiment at NICA and the upgraded NA61 at CERN. All these experiments will explore collisions at low energies for which the associated baryochemical potentials are large, and the antibaryons are much suppressed with respect to baryons. Therefore, the dynamical effects are supposed to be dominated by the latter, like nucleons.

From the idea of measuring (net-)proton correlations we have focused our attention to the possible modifications of their interaction (the $NN$ potential) under the influence of the QCD CP. The main idea of this work is the following: the attractive part of the nuclear potential at inter-particle distances around $r=1$ fm is dominated by the $\sigma$ exchange, the excitation of the critical mode. When these excitations become very light (ideally massless) close to the CP, the nucleons will experience a stronger attraction of longer range (of the order of $1/m_\sigma$). Therefore important correlations between several nucleons, encoded in the higher-order cumulants, would build up close to $T_c$. Even more, if the hadronic evolution spends enough time in the vicinity of critical region, this attraction would be able to bound several nucleons and form light nuclei like $^3$H, $^3$He or $^4$He. It would be very remarkable if nuclear physics is able to guide us in the search of the QCD CP.

\section{Critical mode and $NN$ interaction}

\begin{figure}
\includegraphics[width=.5\textwidth]{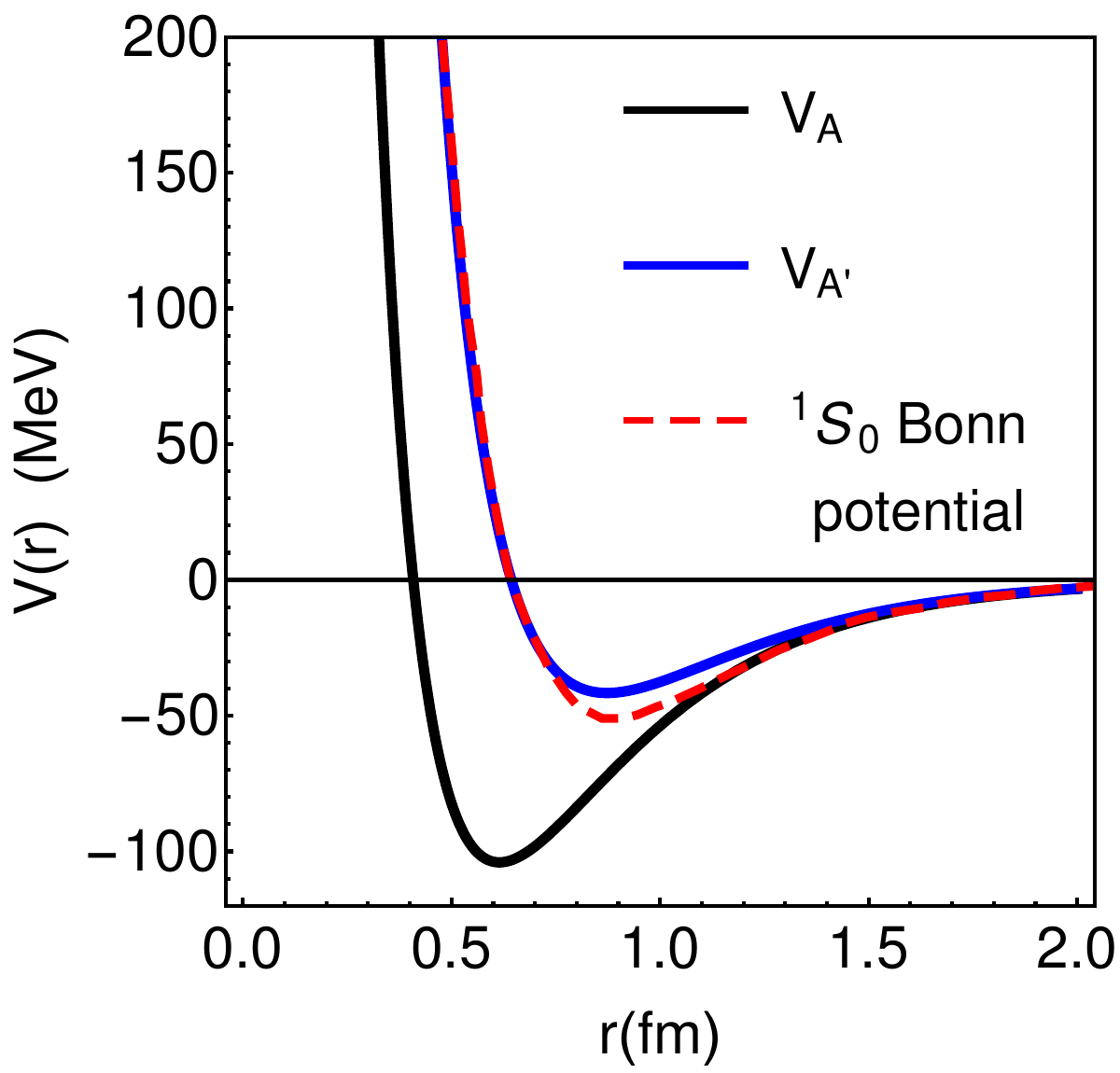}
\includegraphics[width=.5\textwidth]{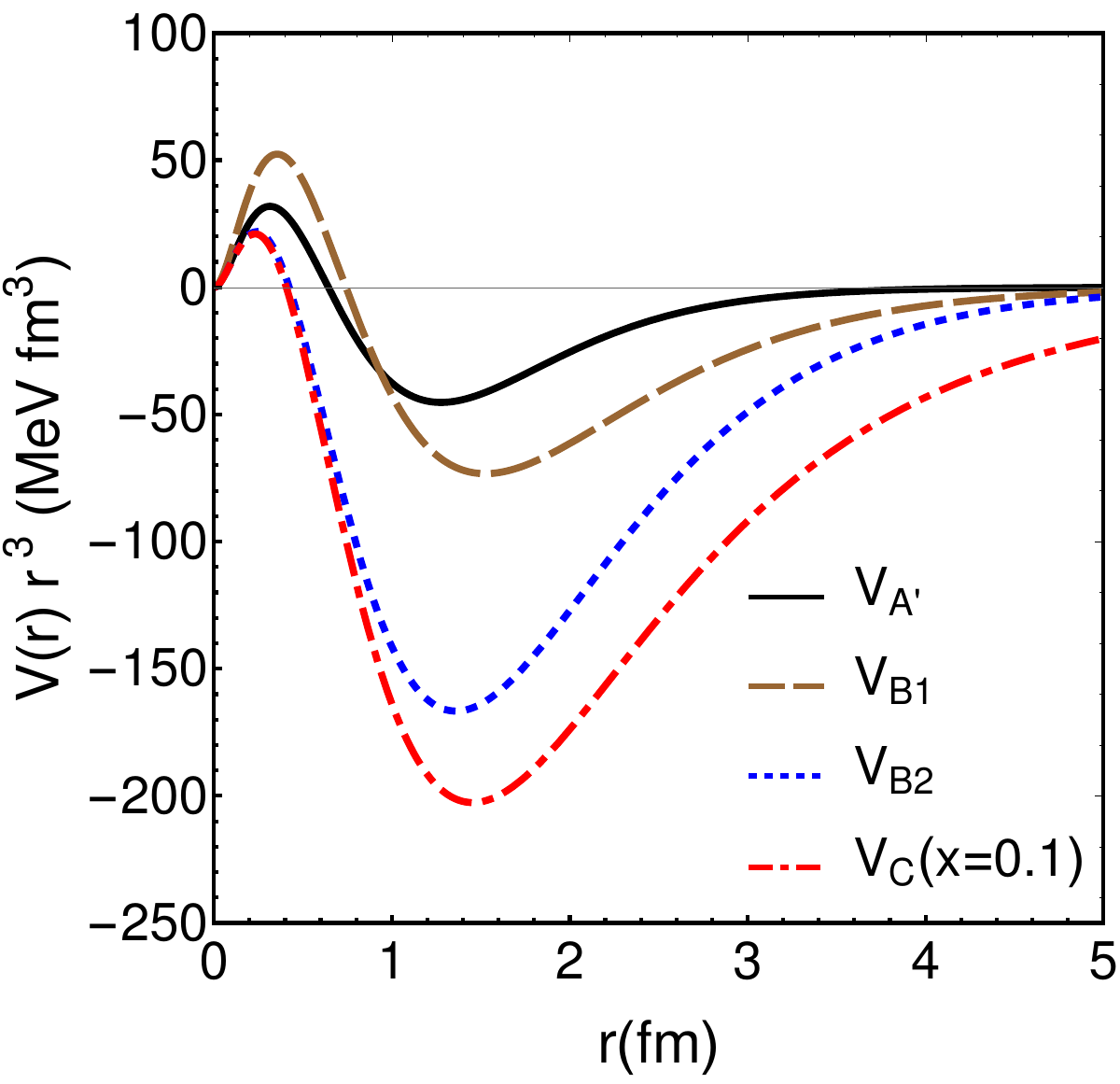}
\caption{Left panel: $NN$ potential in the original Serot-Walecka model $(V_{A})$~\cite{Serot:1984ey}; with a 40\% increased repulsion $(V_{A'})$~\cite{Shuryak:2018lgd}; and the Bonn potential~\cite{Machleidt:2000ge}. Right panel: Different versions of the $NN$ potential (scaled by $r^3$) in the vicinity of the QCD CP, where the mass of the $\sigma$ mode is reduced. The definitions of these potentials are given in Eqs.~(\ref{eq:pot1},\ref{eq:pot2},\ref{eq:pot3},\ref{eq:pot4}).}
\label{fig:Walecka_Bonn}
\end{figure}

The main properties of the $NN$ potential can be described with the simple Serot-Walecka model~\cite{Serot:1984ey}. In this model the nuclear interaction is described in terms of isoscalar mesons exchanges. At short inter-nucleon distances the repulsion forbidding collapse is mediated by the vector $\omega$ meson. At large distances the attraction due to the $\sigma$ mode allows nuclear matter to be bound. The $NN$ potential in this model reads, 
\begin{equation} 
V_A(r)=- \frac{g_\sigma^2}{4\pi r}e^{-m_\sigma r}+ \frac{g_\omega^2}{4\pi r}e^{-m_\omega r} \ , \label{eq:VA} 
\end{equation}
where $r$ is the inter-nucleon distance, and the label $A$ denotes the use of parameters fixed at mean field in~\cite{Serot:1984ey}. These read $m_N=938$ MeV, $m_\sigma=500$ MeV, $g_\sigma^2 = 267.1 m_\sigma^2/m_N^2$, $m_\omega=782$ MeV, and $g_\omega^2 = 195.9 m_\omega^2/m_N^2$. The $NN$ potential presents a minimum at $r \sim 0.6$ fm, and it is shown in a black solid line in the left panel of Fig.~\ref{fig:Walecka_Bonn}.

Our initial potential $V_{A'}$~\cite{Shuryak:2018lgd} has the same form of Eq.~(\ref{eq:VA}), but we do not attach to the mean-field parameters, and increase the $g^2_\omega$ coupling by 40\% so that the resulting potential is more shallow and closer to the Bonn potential~\cite{Machleidt:2000ge} (see Fig.~\ref{fig:Walecka_Bonn}). Notice that when working beyond mean field one does not require a very deep potential like $V_{A}$, because many-body dynamics will generate the necessary correlations to bound nuclear matter.

The main idea in our analysis is that the properties of the medium formed at HICs might alter the parameters of this potential, especially close to $T_c$ (the critical temperature). In particular, the critical mode $\sigma$~\cite{Stephanov:1998dy,Stephanov:1999zu} suffers strong modifications and its mass becomes small close to $T_c$, 
\begin{equation} m_\sigma \sim \frac{1}{\xi} \sim \left( \frac{|T-T_c|}{T_c} \right)^{\nu} \ , \end{equation}
where $\xi$ is the correlation length of the critical mode. 

The implications of this mass reduction in the $NN$ potential are crucial. The attraction between nucleons gets enhanced and long ranged (to distances of the order $\xi$). This attraction is not compensated by a similar increase of the repulsion, and the precise cancellation between them in cold nuclear matter does not hold anymore. In a realistic system (with finite boundaries and a limited influence of the critical dynamics) the mass cannot go all the way to zero, and we should only expect a moderate reduction. In the first study performed in~\cite{Shuryak:2018lgd} we examine several potentials, where the $\sigma$ mass is considered at most, a factor $\sqrt{6}$ less than its vacuum value. 

In addition to the potential $V_A$ (with mean-field parameters) and $V_{A'}$ (with increased repulsion) we will consider 3 more versions, each one with more degree of criticality: $V_{B_1}$ is obtained from $V_{A'}$ by reducing both $m_\sigma^2$ and $g^2_\sigma$ a factor 2;
$V_{B_2}$ from $V_{A'}$ by decreasing the $m_\sigma^2$ a factor 2 but keeping the same coupling; and finally, a 1-parameter potential $V_C(x)$ which interpolates between $V_{B_2}$ and one with a very light critical mode $m^2_\sigma \rightarrow m_\sigma^2/6$,

\begin{eqnarray}
V_{A'}  &=& V_{A} (g_\omega^2 \rightarrow 1.4 g_\omega^2) \ , \label{eq:pot1} \\
V_{B_1}  &=& V_{A'} (m_\sigma^2 \rightarrow m_\sigma^2/2 ; g_\sigma^2 \rightarrow g_\sigma^2/2) \ , \label{eq:pot2} \\
V_{B_2}  &=& V_{A'} (m_\sigma^2 \rightarrow m_\sigma^2/2) \ , \label{eq:pot3} \\
V_C (x) &=& (1-x) V_{B_2} + x \ V_{A'} (m^2_\sigma \rightarrow m^2_\sigma/6) \quad x\in(0,1) \ . \label{eq:pot4}
\end{eqnarray}

These potentials (multiplied by $r^3$) are plotted in the right panel of Fig.~\ref{fig:Walecka_Bonn}. The increasing attraction of the $NN$ interaction is evident as long as the critical dynamics is dominating more and more ($V_{A'} \rightarrow V_{B_1} \rightarrow V_{B_2} \rightarrow V_{C}$).

A simple preliminar calculation in a mean-field approach gives the binding energy of a nuclear drop as a function of its size and the $NN$ potential used. The results can be seen in~\cite{Shuryak:2018lgd}, and they simply confirm the expected effect: the ``noncritical'' potentials $V_A, V_{A'},V_{B_1}$ cannot bound nuclear matter of smaller size, and the bigger ones are only slightly bound. Only the most critical potentials are able to hold these nuclear drops, even the smallest ones (cf. Fig. 3 in~\cite{Shuryak:2018lgd}). Nevertheless, this calculation is not fully consistent because such a strong attraction would generate two-body correlations, which are neglected at mean-field level. Therefore, we need to consider a many-body approach able to describe nuclear  correlations.

For this goal we solve a classical nonrelativistic Molecular Dynamics scheme~\cite{Gelman:2006sr} with a finite number of nucleons interacting through a pairwise potential. The temperature of the system is fixed by the light degrees of freedom (thermal bath), which we encode in a Langevin dynamics. Therefore, in the equations of motion we include a stochastic force for the nucleons as well as a drag force $\lambda$, proportional and opposed to the nucleon momentum, 
\begin{equation} \left\{
\begin{array}{rcl}
 \dfrac{ d \vec x_i}{dt}   & = & \dfrac{\vec p_i}{m_N} \ , \\
  \\
 \dfrac{ d \vec p_i}{dt} & = &   - \sum\limits_{j\neq i} \dfrac{ \partial V (|\vec x_i- \vec x_j|)}{\partial \vec x_i} -\lambda \vec p_i +  \vec \xi_i \ , 
\end{array}
\right.
\end{equation}
where $i=1,..,N$, and $\vec \xi$ is the random noise following a white Gaussian distribution,	
\begin{eqnarray}
  \langle \vec \xi_i (t) \rangle  &=& 0 \ , \\
 \langle \xi^a_i (t) \xi^b_j (t') \rangle  &=& 2T\lambda m_N \delta^{ab} \delta_{ij} \delta(t-t') \ , 
 \end{eqnarray}
with $a,b=1,2,3$. Making use of the fluctuation-dissipation theorem we relate $\lambda$ with the variance of the noise. In~\cite{Shuryak:2018lgd} we used $\lambda=0.256$ fm$^{-1}$.

\begin{figure}
\includegraphics[width=.5\textwidth]{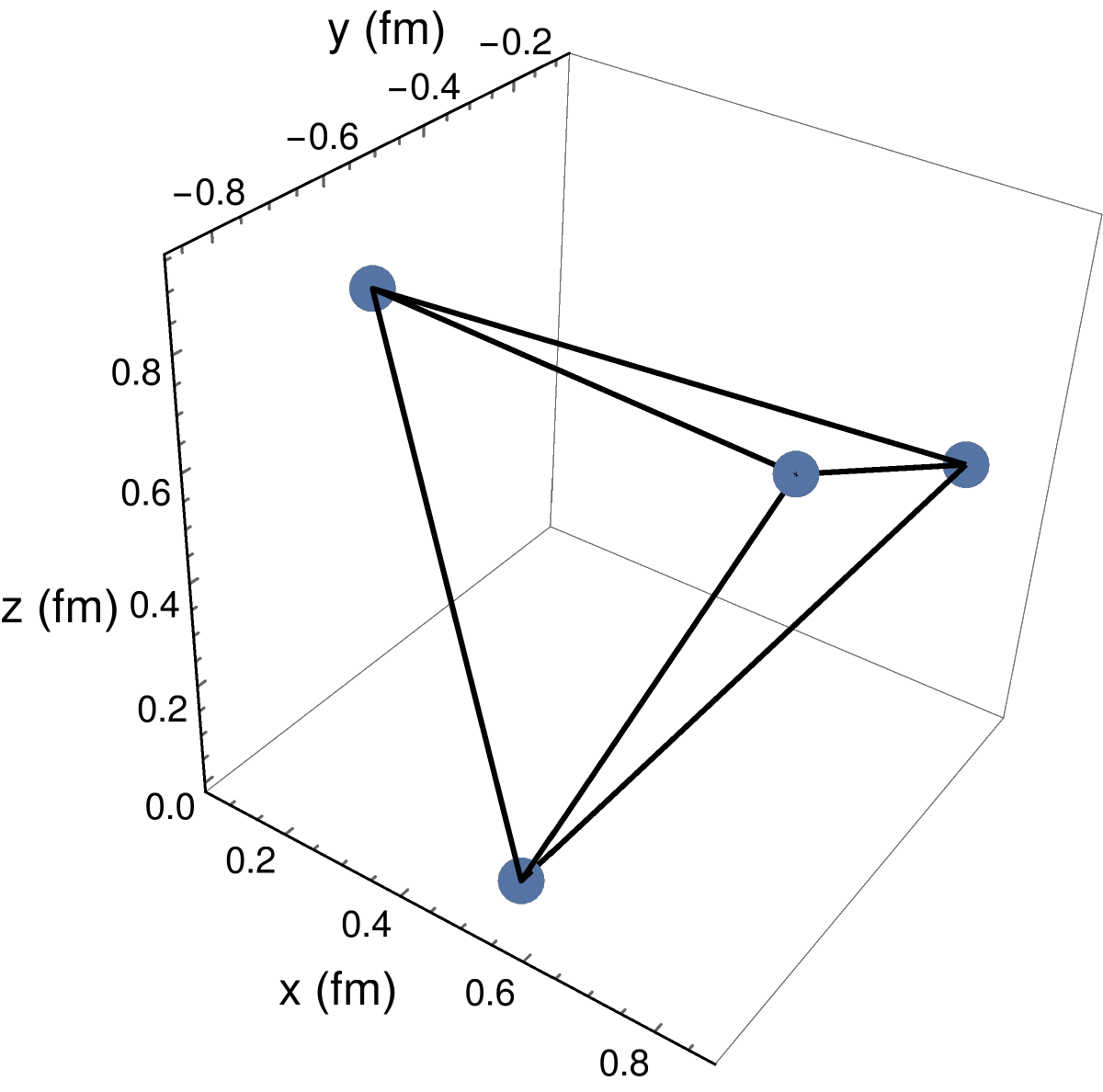}
\includegraphics[width=.5\textwidth]{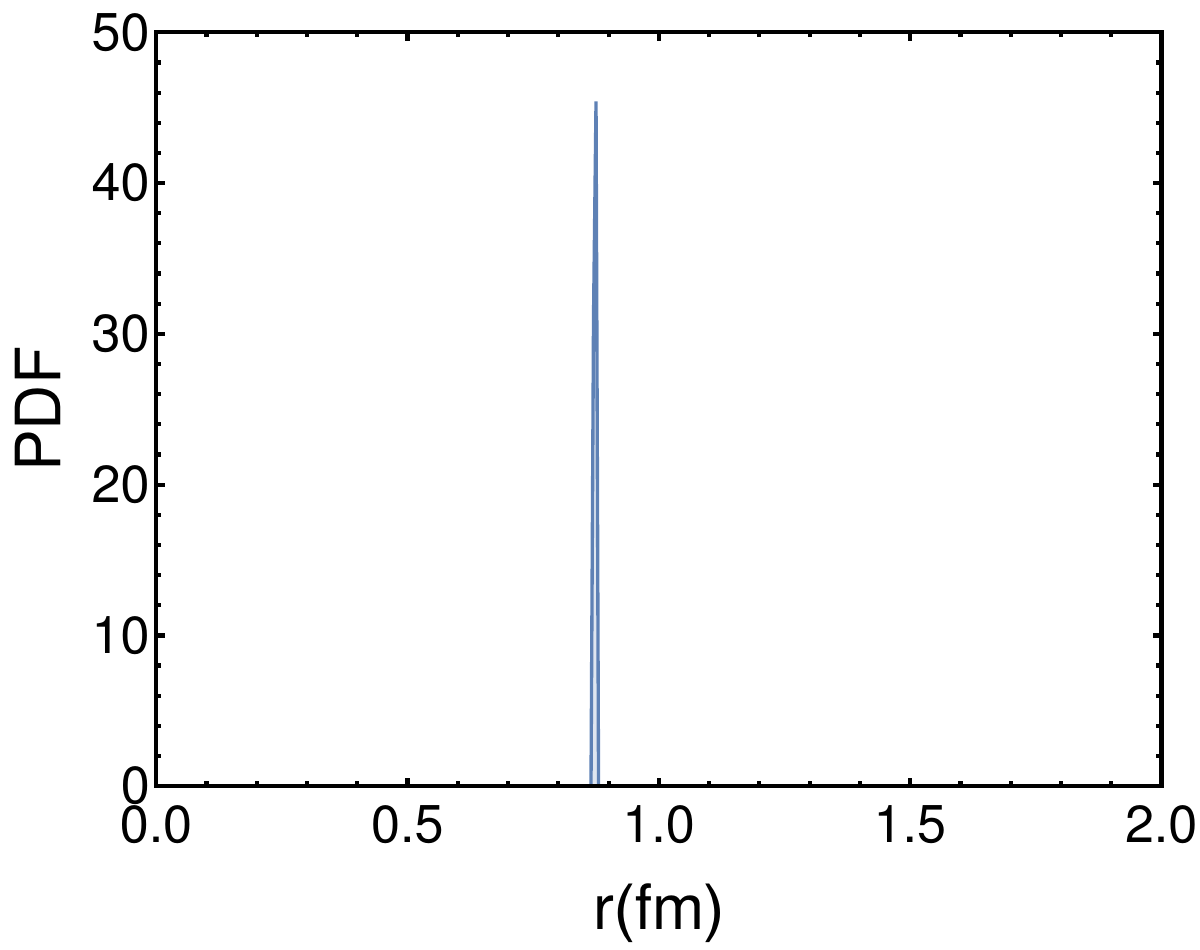}

\caption{Left panel: Cold ($T=10^{-3}$ MeV) configuration for $N=4$ nucleons interacting through the potential $V_{A'}$. Right panel: Probability distribution function of the inter-particle distances. As expected~\cite{Shuryak:2018lgd}, the single peak is located at the minimum of the pairwise potential.}
\label{fig:N4}
\end{figure}

\begin{figure}
\includegraphics[width=.5\textwidth]{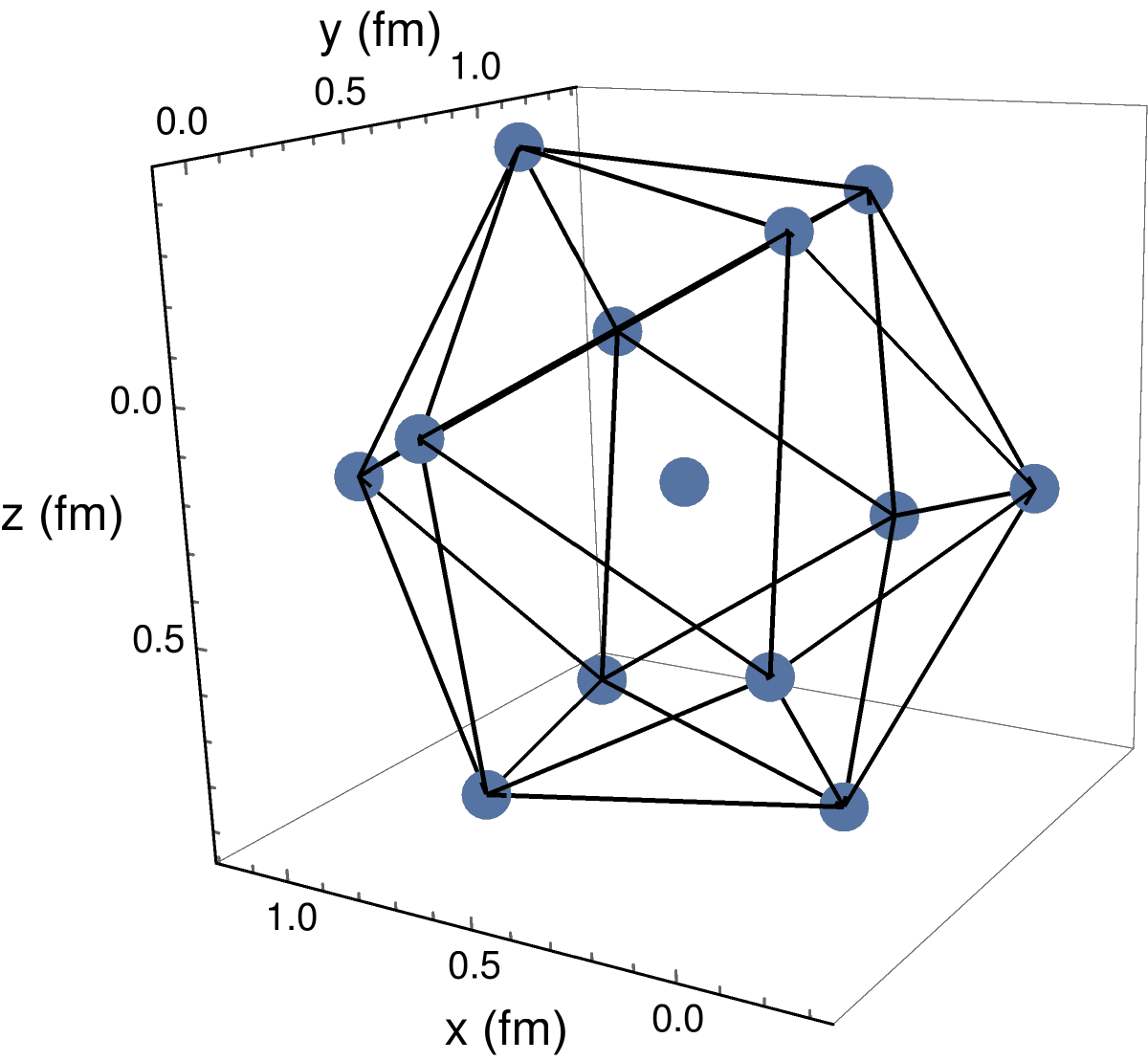}
\includegraphics[width=.5\textwidth]{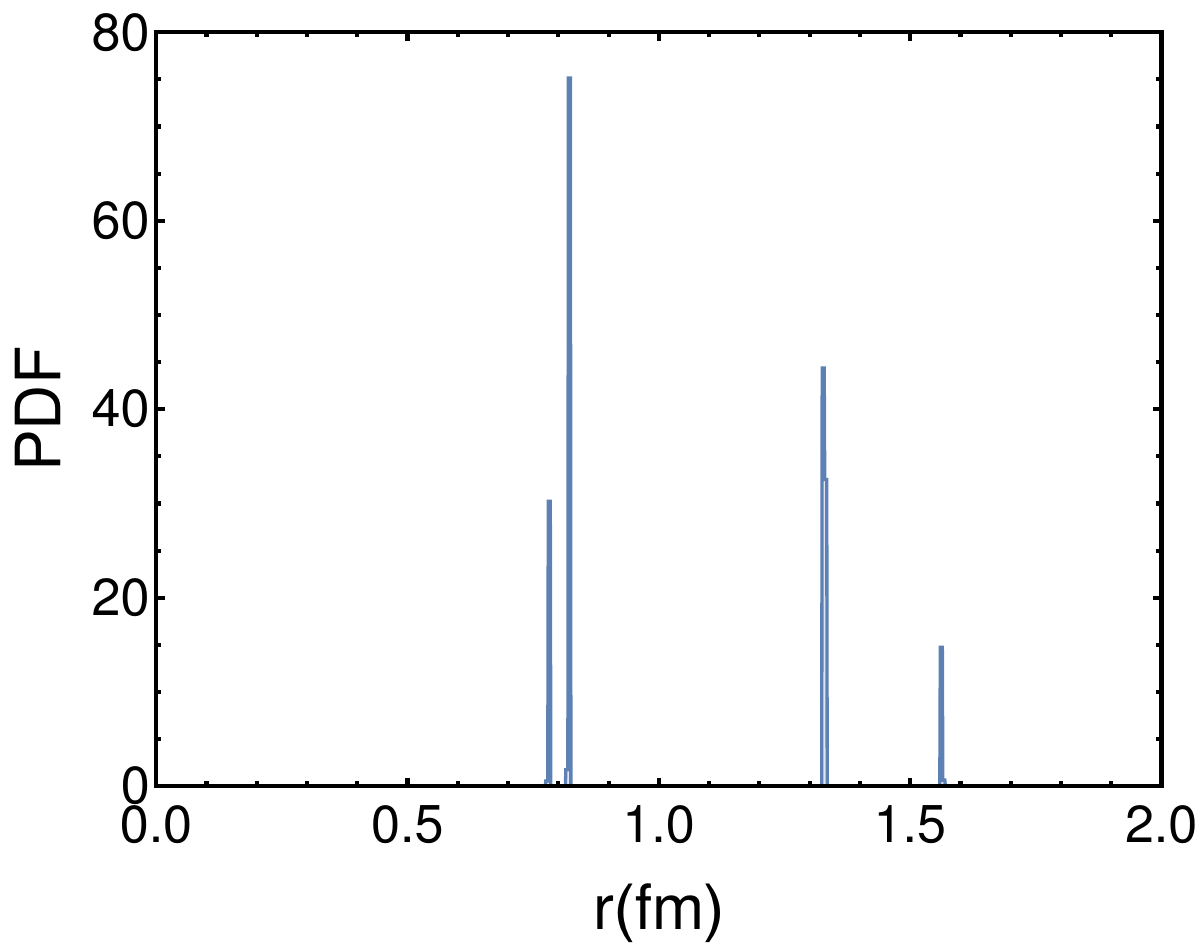}
\caption{Left panel: Cold ($T=10^{-3}$ MeV) configuration for $N=13$ nucleons interacting via potential $V_{A'}$. Right panel: Probability distribution function of the inter-particle distances. As expected~\cite{Shuryak:2018lgd}, one finds 4 peaks at distances with ratios $1:\sqrt{2-2/\sqrt{5}}:\sqrt{2+2/\sqrt{5}}:2$.}
\label{fig:N13}
\end{figure}

Examples of cold configurations ($T=10^{-3}$ MeV) with few nucleons can serve to test the numerical routine and check well-known expectations from symmetry arguments. For example, for $N=4$ and $N=12$ the dynamics places the nucleons at the vertices of Platonic solids, viz. tetrahedron and icosahedron. The analysis of the relative distances and angles confirm these shapes. In the left panel of Fig.~\ref{fig:N4} we show a snapshot of the configuration of $N=4$ nucleons at some time after thermalization under the influence of $V_{A'}$. 
In the right panel, a probability distribution function of the mutual distances shows the consistency with the tetrahedral configuration. The single peak of the distribution coincides exactly with the minimum of the potential. In Fig.~\ref{fig:N13} we repeat the calculation for a medium size system $N=13$ at $T=10^{-3}$ MeV and the same potential. The probability distribution function of distances matches exactly the icosahedral distribution (+1 particle in the center). Small temperatures still preserve the geometrical shapes but broaden the peaks of the distributions, due to the thermal motion of the nucleons.

\begin{figure}
\includegraphics[width=.5\textwidth]{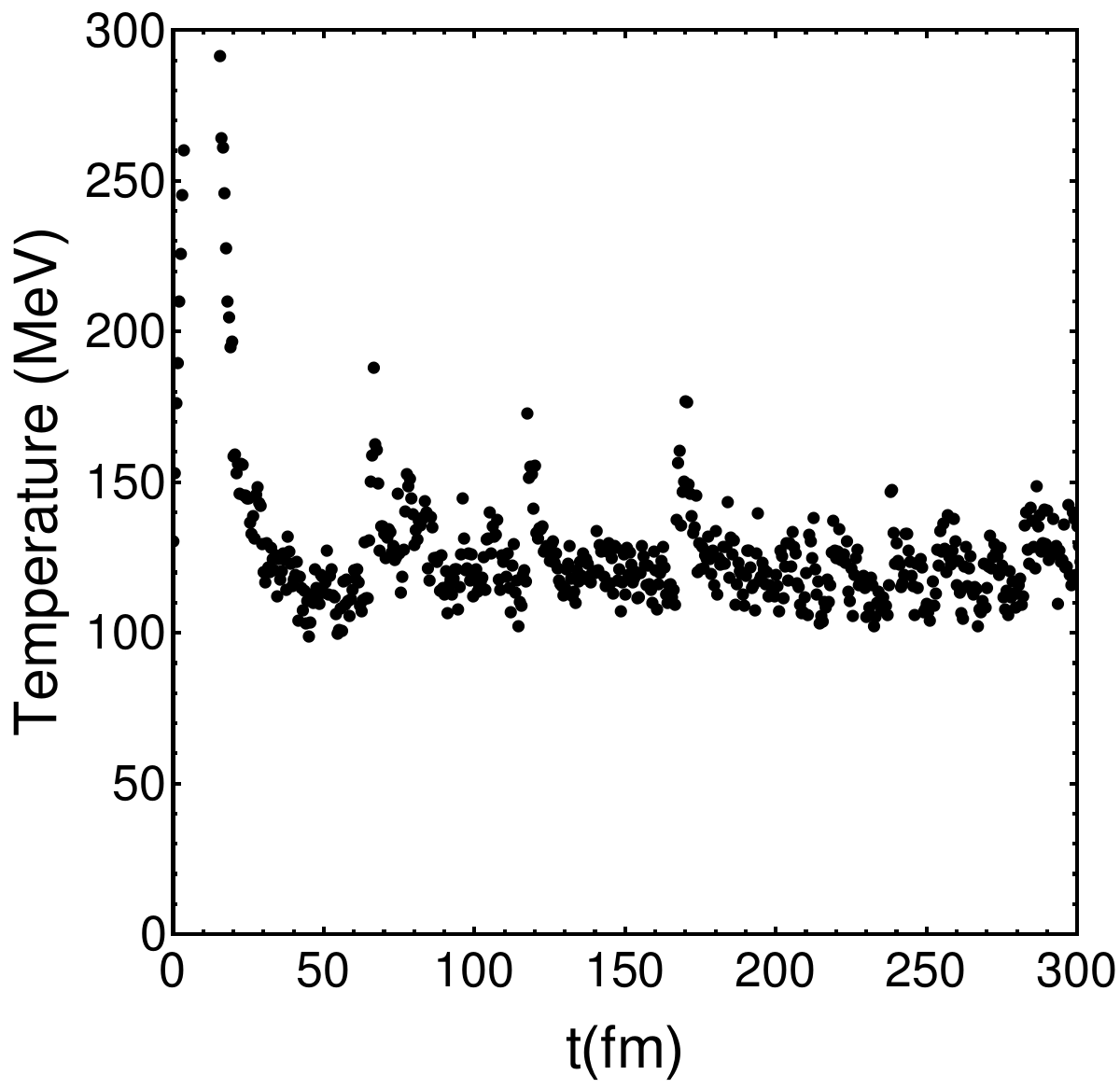}
\includegraphics[width=.5\textwidth]{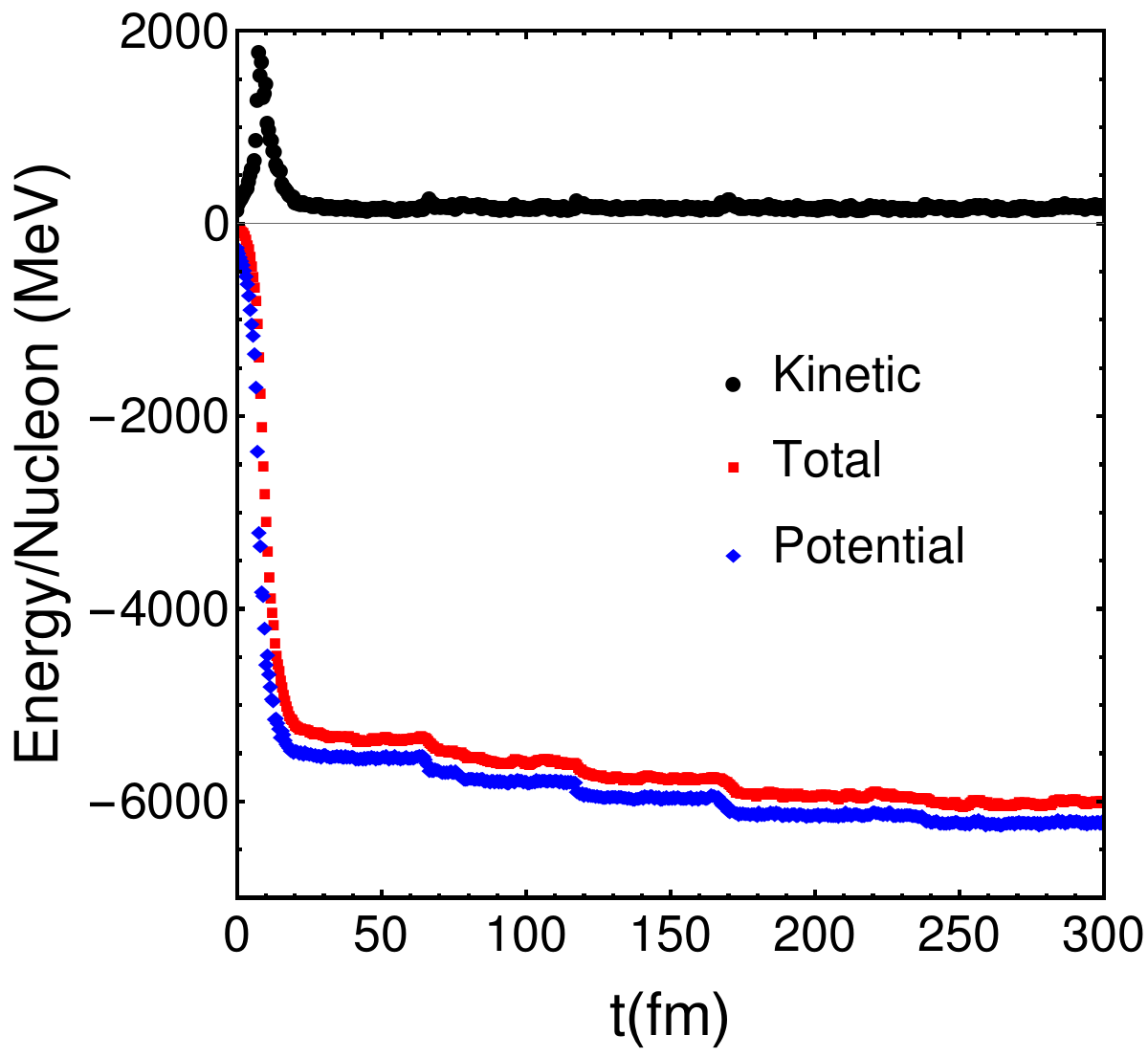}
\includegraphics[width=.5\textwidth]{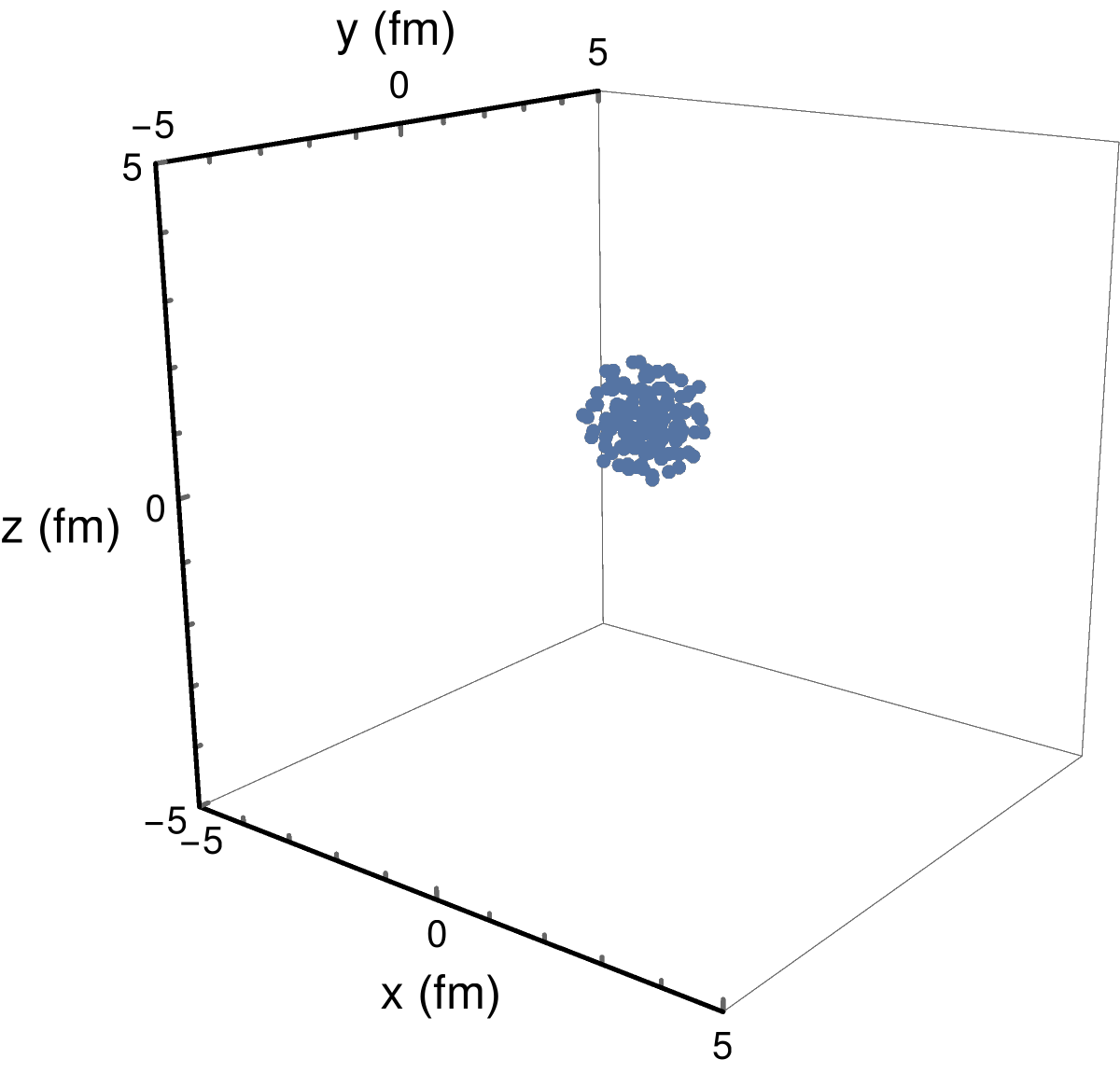}
\includegraphics[width=.5\textwidth]{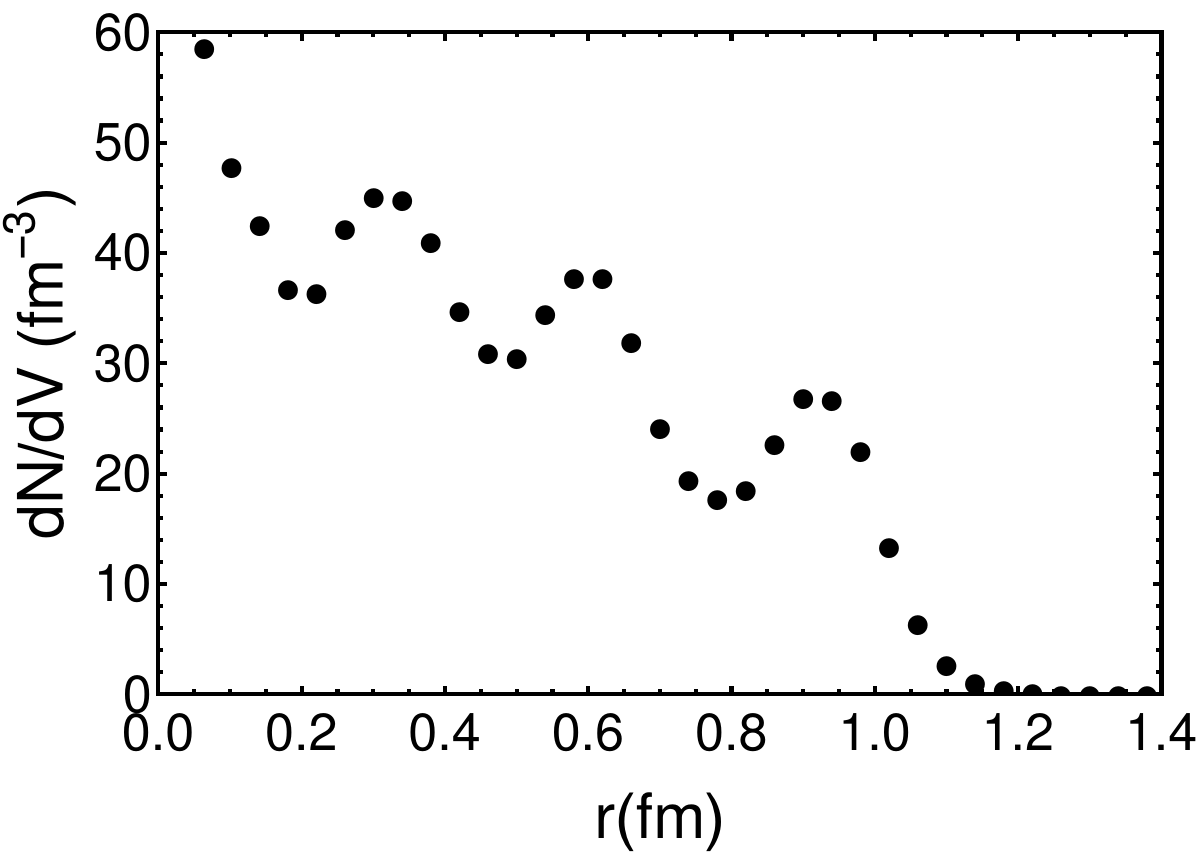}
\caption{Results of a simulation with $N=128$ nucleons at $T=120$ MeV. Top-left panel: ``Temperature'' of the system versus time. Top-right panel: Potential, kinetic and total energies per nucleon as a function of time. Bottom-left panel: Configuration of nucleons after thermalization. Bottom-right panel: Internal distribution of nucleons $dN/dV$ measured from the centroid of the cluster.}
\label{fig:N128}
\end{figure}

To approach the physical scenario we run a large system with $N=128$ nucleons at a temperature close to freeze-out $T=120$ MeV. This is a strongly correlated system where the nuclear potential is felt by all 8256 mutual distances. The combined effect of all these interactions produces nuclear clustering after thermalization. In Fig.~\ref{fig:N128} we observe a summary of the results for this many-body case. In the top left panel we present the ``temperature'' of the system versus time. The ``temperature'' is a measure of the average kinetic energy per particle multiplied by 2/3, so that in equilibrium it would correspond to the true temperature ($T=120$ MeV). In the top right panel we plot the kinetic, potential and total energies per nucleon versus time. Notice that the system is dissipative and the total energy is not conserved. In addition, note the huge potential energy developed in the system. In the bottom left panel we show the 3D configuration of the system, for some particular time after equilibration. Clustering is evident. Finally, the internal structure of the cluster $dN/dV$ is shown in the bottom right panel. It resembles a shell-like structure with peaks at regular distances from the center of the cluster. This explicitly shows that strong correlations are generated between nucleons and a mean-field approach would be simply inadequate.

\section{Higher-order (net-)proton moments and cumulants}

We are ready to apply our model to a system simulating heavy-ion collision at BES energies as measured by STAR collaboration~\cite{Adamczyk:2017iwn}. We will consider two different kinematic cuts which have been applied in the analyses of these data. We will denote {\it Cut 1} as the one with rapidity $|y|<0.5$ and $0.5 \textrm{ GeV}/c < p_\perp < 0.8 \textrm{ GeV}/c$~\cite{Adamczyk:2013dal}; whereas {\it Cut 2} has the same rapidity range but extends the $p_\perp$ coverage up to 2 GeV$/c$~\cite{Luo:2015ewa}.

To mimic the conditions of the BES as measured by STAR we set a calculation with $N=32$ nucleons in a medium at temperature $T=150$ MeV (average temperature between hadronization and freeze-out) with a baryonic density of $n=0.3$ fm$^{-3}$~\cite{Adamczyk:2017iwn,Ivanov:2018vpw}. The duration of the simulation is set to $\Delta t=5$ fm, which is a conservative time for low-energy collisions. We do not include the effects of the fireball expansion in the evolution, but we perform a final mapping of the kinematic variables to fit the experimental $p_\perp$ and a flat distribution in rapidity. We repeat the simulation a number of events $10^5$ to achieve similar statistics as in experiment. For more details on these numbers and their justification we refer the reader to our publication~\cite{Shuryak:2018lgd}. We use the results at $\sqrt{s_{NN}}=19.6$ GeV as a baseline for a non-critical scenario. In our simulation this is achieved by the potential $V_{A'}$ (no $\sigma$-mass modifications).

The only parameter which is not fixed a priori is $N$, as experimentally we only know the average number of protons in a given kinematic cut. Focusing on the {\it Cut 1} we compare our value of $C_1$ (average number of protons) and compare it to the experimental result. Then, we use the ratio between the two to rescale all our proton moments by the same amount. Once this is done, we are able to generate all other moments for both {\it Cut 1} and {\it Cut 2}.

\begin{figure}
\includegraphics[width=.5\textwidth]{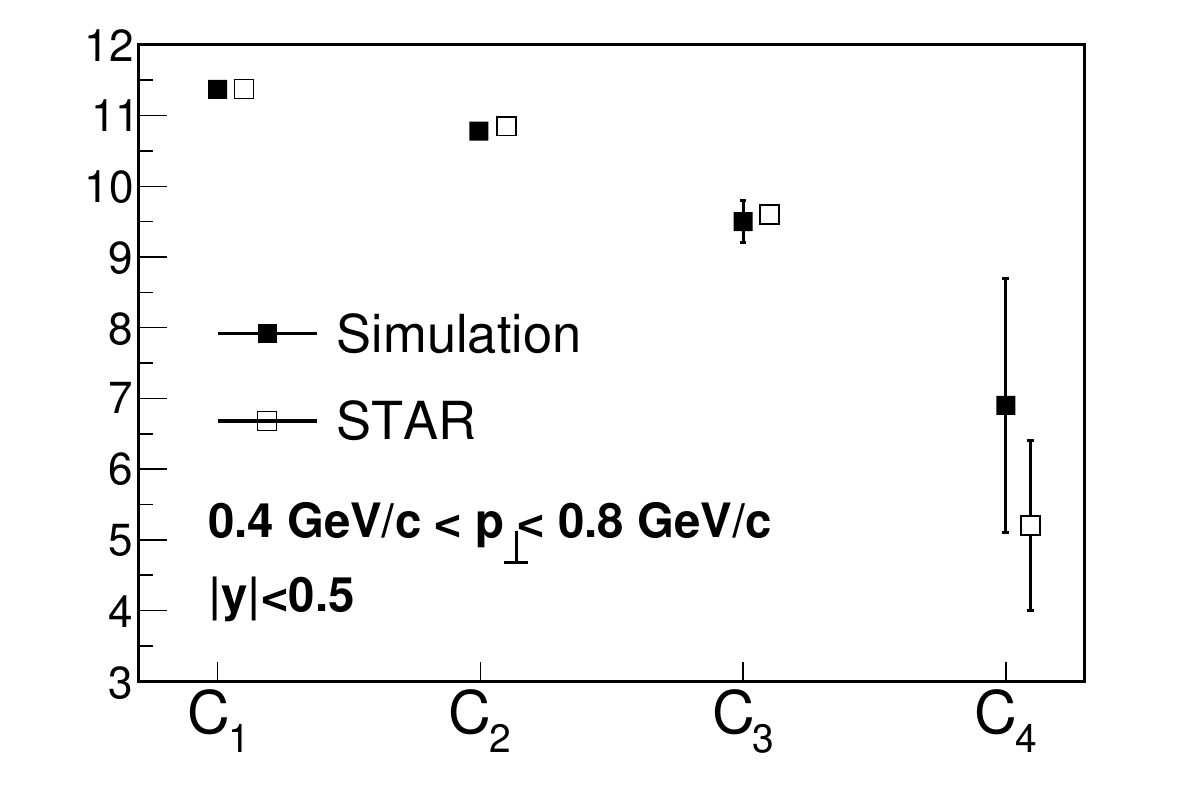}
\includegraphics[width=.5\textwidth]{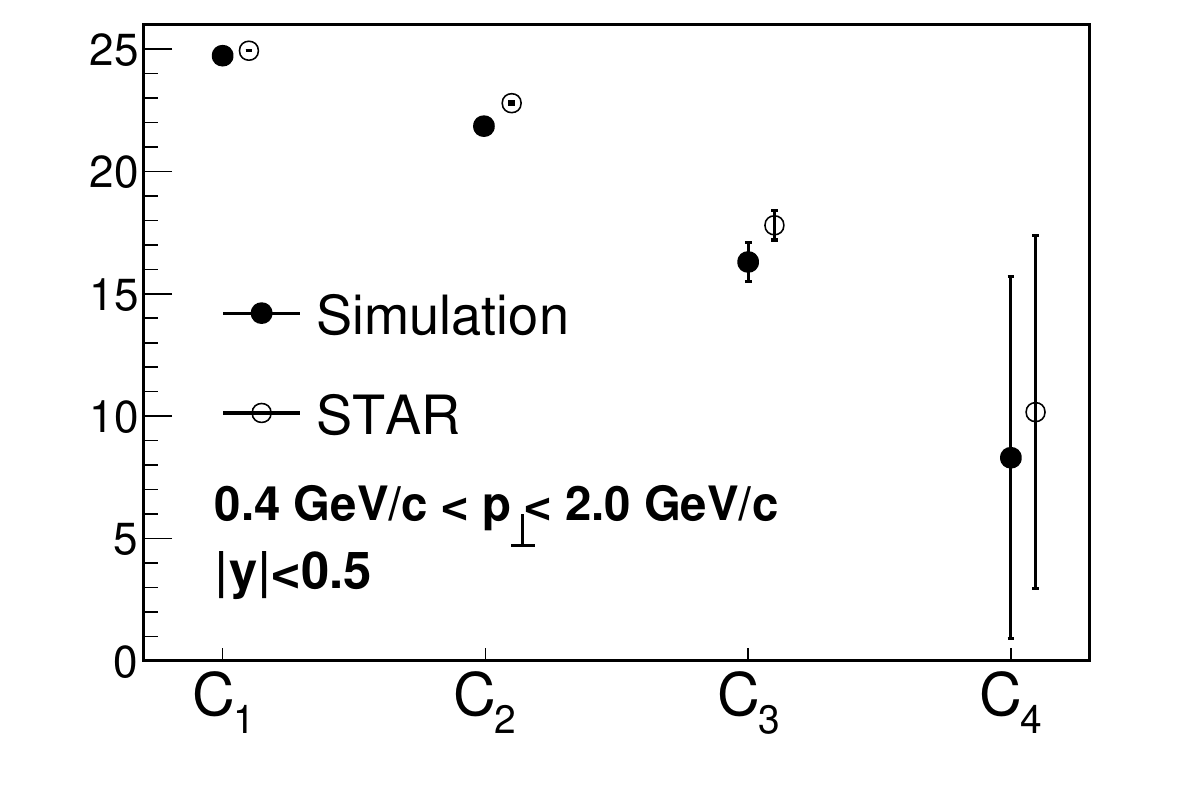}
\caption{Moments of the proton distribution obtained in our simulation with a non-critical potential $V_{A'}$ compared to the experimental results of STAR at a collision energy of $\sqrt{s_{NN}}=19.6$ GeV. Left and right panels show the result for the {\it Cut 1} and {\it Cut 2}, respectively.}
\label{fig:Corr}
\end{figure}

For this particular energy $\sqrt{s_{NN}}=19.6$ GeV (where no critical dynamics are expected), the results are summarized in Fig.~\ref{fig:Corr}. We observe a reasonable agreement (both in the moments and their error bars) between the experimental data and our simulations with the noncritical potential $V_{A'}$.

Using the cumulants we can finally compute the scaled skewness ($S\sigma$) and kurtosis ($\kappa \sigma^2$), defined by 
\begin{equation} S\sigma = \frac{C_3}{C_2} \quad , \quad \kappa \sigma^2 = \frac{C_4}{C_2} \ ,  \end{equation}
and repeat our simulations using different potentials $V_{B_1},V_{B_2},V_{C}$. To isolate the effect of the interaction potential, we do not modify any other parameter in the simulation. The rationale behind this exercise is that once the proton moments for the collision energy $\sqrt{s_{NN}}=19.6$ GeV are compatible with Poissonian fluctuation (no critical dynamics), there should be lower energy for which the system evolves close to the critical point. As long as one approaches that collision energy, the $NN$ potential becomes more and more critical. We want to study how the higher-order cumulants are continuously modified as long as we approach that energy. Unfortunately we cannot match each potential with a corresponding collision energy without more modeling.

\begin{figure}
\includegraphics[width=.5\textwidth]{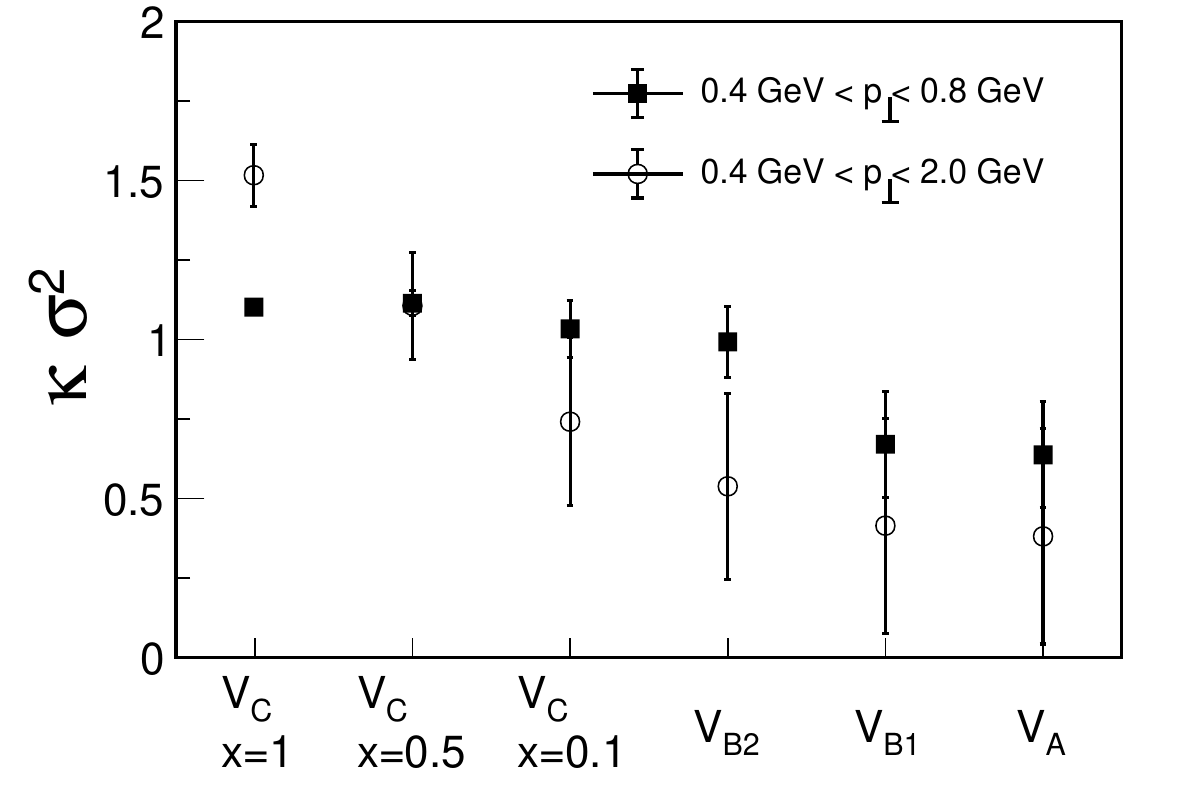}
\includegraphics[width=.5\textwidth]{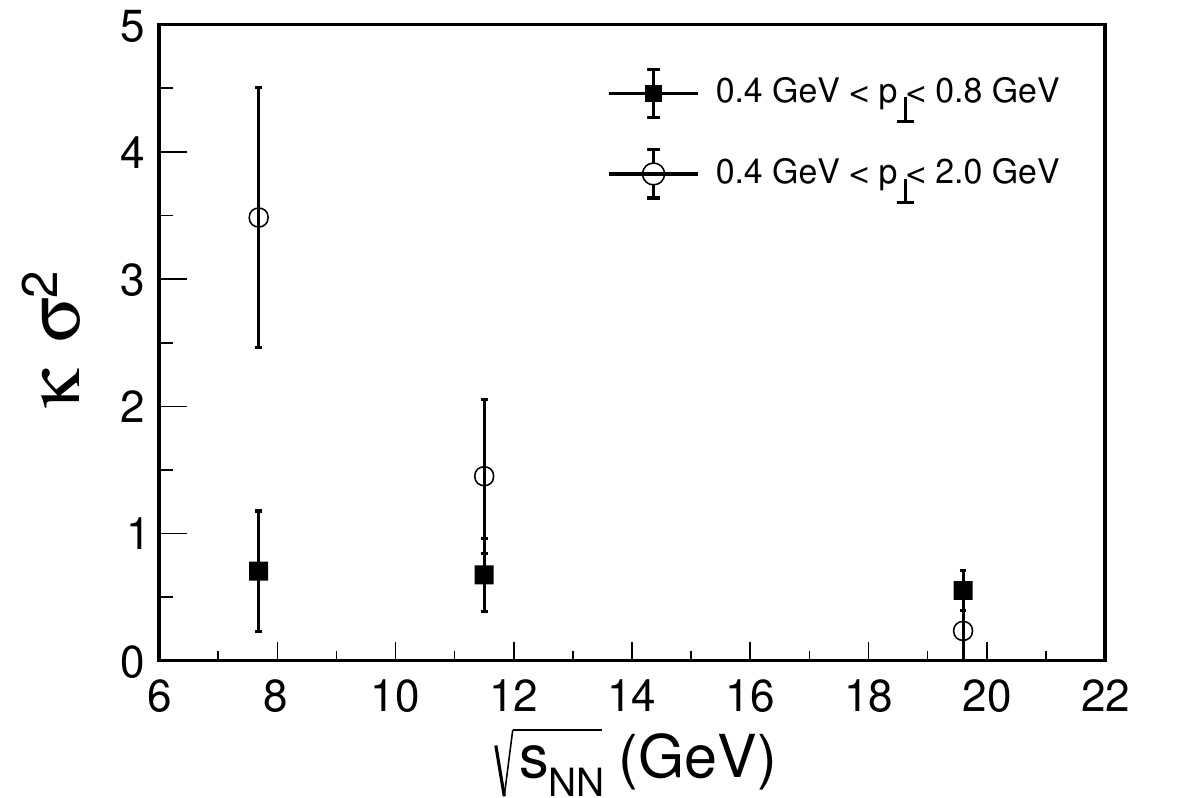}
\caption{Left panel: Theoretical scaled kurtosis as a function of the $NN$ potential. The critical enhancement increases from right to left in the OX axis. Right panel: Experimental scaled kurtosis from STAR measurements as a function of the collision energy.}
\label{fig:Kur}
\end{figure}

The results for the scaled kurtosis are shown in Fig.~\ref{fig:Kur}. In the left panel we present our results as functions of the $NN$ potential (we approach the critical region going from right to left along the OX axis). In the right panel we plot the experimental data from STAR at the lowest collision energies. The numbers obtained in our simulations are realistic in spite of the crude model used. In both panels, notice that for the {\it Cut 1} (solid symbols) the increase of $\kappa \sigma^2$ is very mild, but in the {\it Cut 2} (open symbols) there is an increase of several units. Our conclusion is that (at least part of) the increase of the scaled kurtosis seen in experimental data is compatible with the effect of $NN$ potential modification close to $T_c$~\cite{Bzdak:2016jxo}. 

\section{Light nuclei formation at the CP}

If the fireball spends enough time in the vicinity of $T_c$ it will be possible for the attractive $NN$ potential to bind nucleons and form nuclear clusters. While the statistical thermal model works very nicely at high energies for particles as heavy as $^4$He~\cite{Andronic:2017pug}, the presence of the QCD CP would increase the multiplicity of light nuclei due to nuclear clustering. 

\begin{figure}
\begin{center}
\includegraphics[width=.5\textwidth]{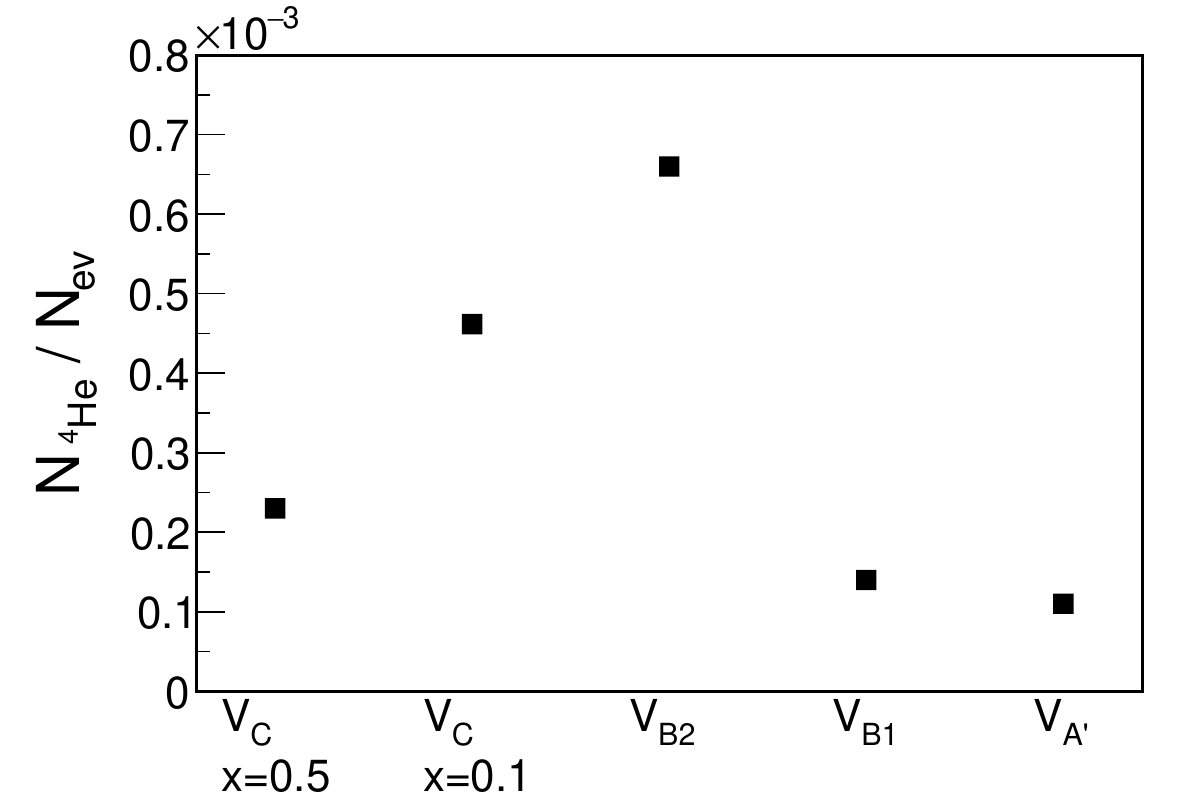}
\end{center}
\caption{Number of 4-nucleon clusters per event as a function of the $NN$ potential. The approach to the critical region is understood as going from right to left along the OX axis. For the most attractive potentials the number of 4-nucleon clusters actually decreases, because the nucleons are in fact contained in bigger clusters.}
\label{fig:pre4He}
\end{figure}

In our simulation we look for clusters of 4 nucleons close in phase space, which are understood as potential candidates for $^4$He nuclei. We have scanned the final state for isolated sets of 4 nucleons with inter-particle distance $\Delta r <2$ fm and $\Delta p<0.22$ GeV (for each momentum component). The number of 4-nucleon clusters per event is plotted in Fig.~\ref{fig:pre4He} as a function of the $NN$ potential. This number increases with the attraction (criticality) of the potential, up to the point in which the attraction becomes so large for $V_C$ that nucleons start forming part of bigger clusters, in fact, producing a decrease.

The immediate global observable one would consider to test this prediction is the light-nuclei yield with respect to the statistical thermal expectations~\cite{Andronic:2017pug} as a function of the collision energy. A maximum of this yield at some energy would indicate a strong nuclear attraction due to the critical point. As this overpopulation of light-nuclei is a tiny fraction of the total yield, it makes more sense to consider ratios of light-nuclei multiplicities. For example, take the ratio~\cite{Sun:2017xrx} $\frac{N_t N_p}{N_d^2}$, which combines the yield of tritons, protons, and deuterons. Applying ideal Boltzmann statistics this ratio produces a trivial result $g=0.29$ coming from numerical factors and the spin and isospin degeneracies,
\begin{equation}
\left. \frac{N_t N_p}{N_d^2} \right|_{ideal\  gas} \simeq g \ . 
\end{equation}
with a cancellation of the nucleon-mass and temperature dependences. 

However if the interaction potential is nonneglibible, in the statistical weight one finds that the Boltzmann factor contains 3 powers of $V$ for triton, but only one power for each deuteron. The ratio should be sensible to a thermal average of the $NN$ potential, 
\begin{equation} 
g^{-1} \frac{N_t N_p}{N_d^2} \sim \langle e^{-V/T} \rangle \ . \label{eq:ratio}
\end{equation}
Close to $T_c$ where the $NN$ potential is considerably deep the typical distances are distributed around the minimum of the potential, where $V<0$. Therefore we predict that this ratio should increase at the critical point.

\begin{figure}
\begin{center}
\includegraphics[width=.7\textwidth]{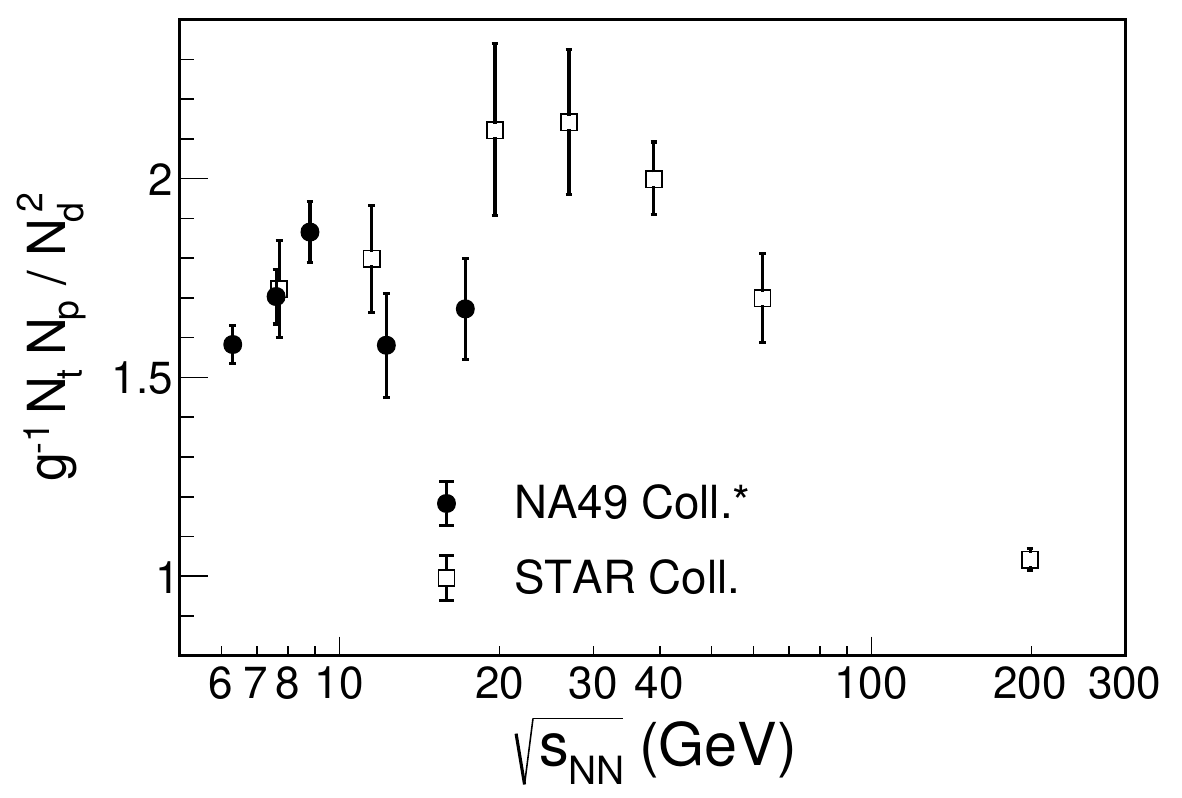}
\end{center}
\caption{Ratio of light nuclei as defined in Eq.~(\ref{eq:ratio}) scaled down by the trivial degeneracy factor $g$. We have adapted the preliminar experimental data from NA49~\cite{Anticic:2016ckv} and STAR collaborations~\cite{Liu:2019ppd}. For the former, we use data collected in~\cite{Sun:2017xrx}.}
\label{fig:Ratio}
\end{figure}

In Fig.~\ref{fig:Ratio} we collect preliminar results from NA49 collaboration~\cite{Anticic:2016ckv} adapted in~\cite{Sun:2017xrx} and STAR experiment~\cite{Liu:2019ppd} and plot them on the same figure. Both collaborations present a ratio which depends on the collision energy with a maximum at some particular energy. Notice that STAR data covers a wider range and the maximum is larger. Also notice that at the highest STAR energy the ratio is compatible with 1, pointing to a situation in which the $NN$ potential is negligible with respect to the temperature. Of course, at that energy the system is known to be close to the crossover transition and far from the possible critical point. In the context of our model it would be interesting to explore different combinations of ratios with extra powers of the nuclear potential for example $N_{^4He} N_p/N_{^3He} N_d \sim \langle e^{-2V/T} \rangle$, or (assuming isospin symmetry) $N_{^4He} N_p^2/N_d^3 \sim \langle e^{-3V/T} \rangle$. In these cases the predicted effect would be more spectacular.

Finally, notice that what we call ``clusters'' are statistical correlation/association of few nucleons which survive the freeze-out. Their energy has large uncertainty (proportional to $T_f$) and they might well decay into $N$ unbound nucleons at the post-freeze-out stage. Experimental evidence for cluster formation should come from the observed multiplicity distribution of light nuclei in low-energy heavy-ion collisions, at much lower temperatures (cf. Fig.~\ref{fig:Ratio}). How many of our clusters can feed down the final light-nuclei yield is the subject of an on-going work.

\section{Conclusions}

In our work~\cite{Shuryak:2018lgd} we have studied the influence of the critical behavior of the $\sigma$ field on the $NN$ potential, and its implications on the dynamics of nucleons in baryon-rich heavy-ion collisions. We have observed that the attractive part of the nuclear potential gets enhanced, and strong nuclear correlations build up during the transit of the fireball close to the QCD CP.

Although the expansion of the system and its limited time evolution tend to diminished this effect, we have obtained that for the experimental conditions of the STAR experiment in the context of the BES program, this nuclear attraction generates an increase at midrapidity of the higher-order proton cumulants, like the scaled kurtosis.

A second implication of the $NN$ potential modification is the increase with respect to the thermal equilibrium of the yields of light nuclei, such as d, t, $^3$He, $^4$He, for the collision energies where the system evolves close to the critical region. While preliminary results from NA49 and STAR collaborations have shown such an increase for some multiplicity ratio involving triton, deuteron and proton yields, here we propose to measure alternative ratios where the effect of the modified potential is larger, for example, involving the $^4$He yield, like $N_{^4He} N_p/N_{^3He} N_d$ or $N_{^4He} N_p^2/N_d^3$.

\acknowledgments

This work was supported in part by the U.S. Department of Energy under Contract No. DE-FG-88ER40388.

\bibliographystyle{JHEP}

\end{document}